\documentclass[10pt,aps,prl,twocolumn,nopacs,final,letterpaper,superscriptaddress,longbibliography]{revtex4-2}

\usepackage{verbatim}  
\usepackage{bm}  
\usepackage{quotes}  
\usepackage{xcolor}  

\usepackage{graphicx}

\usepackage{booktabs}

\usepackage{hyperref} 
\definecolor{goodblue}{RGB}{0, 91, 187}
\hypersetup{
    colorlinks=true,
    linkcolor=goodblue,
    filecolor=goodblue,      
    urlcolor=goodblue,
    citecolor=goodblue,
    pdfborder={0 0 0},
    breaklinks=true,
    pdfpagemode=FullScreen}
\usepackage{parskip}
\usepackage[english]{babel}  

\usepackage{amsmath}  
\usepackage{amssymb}  
\usepackage{amsthm}  
\usepackage{mathtools}  
\usepackage{bbm}  

\usepackage{cleveref}[capitalise]

\makeatletter
\def\maketitle{
\@author@finish
\title@column\titleblock@produce
\suppressfloats[t]}
\makeatother

\renewcommand{\epsilon}{\varepsilon}

\newcommand{\bbN}{\mathbb{N}}
\newcommand{\bbR}{\mathbb{R}}
\newcommand{\boldA}{\bm{A}}

\newcommand{\boldX}{\bm{X}}
\newcommand{\boldx}{\bm{x}}
\newcommand{\boldc}{\bm{c}}

\newcommand{\boldmu}{\bm{\mu}}
\newcommand{\tildec}{\Tilde{c}}
\newcommand{\tildemu}{\Tilde{\mu}}
\newcommand{\ER}{Erd\H os-R\'enyi }

\newcommand{\trans}[1]{{#1}^{\mathrm T}}

\def\NUMNODES{256~}
\def\AVGDEGREE{18~}

\def\NUMSTEPS{1000~}

\begin{document}

\setlength{\parskip}{0pt}
\title{Emergent contagion complexity: Disentangling mechanistic complexity from correlated heterogeneity\vspace{-0.2cm}}

\author{Katerina Tang}
\altaffiliation{These authors contributed equally to this work. Correspondence should be addressed to daniel.kaiser@virginia.edu.}
\affiliation{Center for Applied Mathematics, Cornell University, Ithaca, New York 14853, USA}

\author{Daniel Kaiser}
\altaffiliation{These authors contributed equally to this work. Correspondence should be addressed to daniel.kaiser@virginia.edu.}
\affiliation{Department of Biology, University of Virginia, Charlottesville, Virginia 22903, USA}

\author{William Thompson}
\altaffiliation{These authors contributed equally to this work. Correspondence should be addressed to daniel.kaiser@virginia.edu.}
\affiliation{Vermont Complex Systems Institute, University of Vermont, Burlington, Vermont 05405, USA}

\author{Jean-Gabriel Young}
\affiliation{Vermont Complex Systems Institute, University of Vermont, Burlington, Vermont 05405, USA}
\affiliation{Department of Mathematics \& Statistics, University of Vermont, Burlington, Vermont 05405, USA}

\author{Laurent H\'ebert-Dufresne}
\affiliation{Vermont Complex Systems Institute, University of Vermont, Burlington, Vermont 05405, USA}
\affiliation{Department of Computer Science, University of Vermont, Burlington, Vermont 05405, USA}
\affiliation{Santa Fe Institute, Santa Fe, NM 87501, USA}
\affiliation{Complexity Science Hub Vienna, A-1080 Vienna, Austria}

\author{Nicholas W. Landry}
\affiliation{Department of Biology, University of Virginia, Charlottesville, Virginia 22903, USA}
\affiliation{Vermont Complex Systems Institute, University of Vermont, Burlington, Vermont 05405, USA}
\affiliation{School of Data Science, University of Virginia, Charlottesville, Virginia 22903, USA}

\begin{abstract}
Simple and complex contagions differ mechanistically; multiple exposures act synergistically in the latter but independently in the former.
Yet correlated mixtures of simple contagions may appear complex when inferring global contagion rules, a phenomenon we call "emergent complexity."
We present a measure of contagion complexity and an inferential framework for estimating mixtures of nonparametric contagion rules from time-series data.
Our work reframes past studies on complex contagion by offering heterogeneous mixtures of simple contagions as an alternative explanation.
\end{abstract}

\maketitle

\setlength{\parskip}{4pt}

\paragraph{}\textit{Introduction}---The spread of information, beliefs, and behaviors, together referred to as "social contagion" is often modeled as a complex contagion, in which multiple exposures act synergistically to promote adoption~\cite{valente_network_1996,centola_spread_2010}.
This mechanism is often presented in contrast to simple contagion mechanisms, where one exposure is sufficient for infection and successive exposures independently contribute to the probability of a node becoming infected.
Complex contagions can exhibit qualitatively different dynamics from simple contagions, such as discontinuous transitions \cite{dodds_universal_2004}, superexponential spread \cite{hebert-dufresne_macroscopic_2020}, different responses to modular network structure \cite{nematzadeh_optimal_2014, osullivan_mathematical_2015}, and distinct optimal intervention strategies \cite{guilbeault_topological_2021}.

These differences have motivated a growing literature on inferring the contagion rules underlying observed dynamics.
Existing statistical methods, however, often either assume \textit{a priori} that the transmission mechanism is known~\cite{peixoto_network_2019} or simply seek to explain whether complex or simple contagion rules better explain infection data in synthetic~\cite{cencetti_distinguishing_2023,andres_distinguishing_2025} and empirical~\cite{monsted_evidence_2017,weng_virality_2013} settings.
The simple and complex contagion paradigms can be unified by specifying infection probabilities through a \textit{contagion kernel} mapping the number of infected neighbors to an infection probability, i.e., a dose-response curve~\cite{dodds_universal_2004}.
Recent work introduced a Bayesian inferential method for estimating nonparametric contagion kernels from time-series data~\cite{landry_reconstructing_2024}; however, this study assumed every individual was characterized by the same contagion kernel.

This kernel-based view is useful, but contagion kernel inference is not the same as mechanism selection.
Several models have been shown to map to complex contagion dynamics despite having simple contagion rules: contagions with heterogeneous parameters across space \cite{st-onge_nonlinear_2024}, contagions with memory effects \cite{anttila_mechanistic_2017, st-onge_universal_2021}, or synergistic contagions \cite{hebert-dufresne_macroscopic_2020, hebert-dufresne_one_2025}.
In the first example, the transmission rate of a simple contagion varies across settings, and ignoring this variation leads to inference of a nonlinear contagion kernel \cite{st-onge_nonlinear_2024} because a local increase in incidence not only means more infectious cases but also suggests a higher local transmission rate \cite{hebert-dufresne_simpsons_2026}.
Interacting contagions provide another example: if one infection changes the susceptibility to or transmissibility of another, then unobserved co-infection states make the transmission rate of the observed contagion appear to vary with prevalence, even when the underlying mechanisms are simple \cite{hebert-dufresne_macroscopic_2020}.
Across these studies, the existence of heterogeneous kernels and uncertainty in parameters is the source of what we call \textit{emergent contagion complexity}, reflecting that contagion dynamics can appear governed by a complex kernel when, in reality, they are generated by a mixture of simple (or simpler) contagion kernels.

In this Letter, we distinguish two sources of apparent contagion complexity.
First, a contagion kernel may depart from the independent-exposure form of simple contagions, indicating \textit{mechanistic complexity} in the underlying transmission rule. 
Second, a kernel may appear non-simple because it aggregates over heterogeneous subpopulations with different kernels and structured mixing across the network; we refer to this as \textit{emergent complexity}.
We introduce a score for apparent kernel complexity, use it to characterize sufficient conditions under which correlated heterogeneity makes a global kernel appear complex, and then detail a method by which a complex contagion kernel may be disambiguated from a mixture of simple contagion kernels. 
That is, we aim to distinguish mechanistic complexity from emergent complexity.
Our findings suggest that complexity may be thought of as an emergent property of contagions and help reframe past empirical work on complex contagions.

\paragraph{}\textit{Heterogeneous contagion kernel model and inference}---We generalize the susceptible-infected-susceptible (SIS) contagion process described in Ref.~\cite{landry_reconstructing_2024} to model heterogeneous contagion kernels.
At each time $t,$ a susceptible node $i$ is infected with probability $c_i(\nu),$ where $\nu$ is the number of infected neighbors of node $i$.
An infected node recovers with probability $\gamma.$
Each node's contagion kernel $c_i:\bbN\to[0,1]$ can be represented in nonparametric form as a vector $\boldc_i=\trans{[c_{i,1},\dots,c_{i,N-1}]},$ where $c_{i,\nu}$ denotes node $i$'s probability of infection by $\nu$ infected neighbors and $N$ is the number of nodes in the network.
This flexible representation allows us to consider simple and complex contagion kernels as special cases.
For simple contagions, in which exposures are independent and infect a node with probability $\beta$, $c(\nu)=1-(1-\beta)^\nu$.
For threshold contagion, in which a node adopts deterministically if $\tau$ neighbors have adopted, $c(\nu) = \mathbbm{1}_{\nu \geq \tau}$, where $\mathbbm{1}$ is the indicator function.

To describe the time evolution of this contagion process mathematically, we track the states of all nodes in vector $\boldx(t)=\trans{[x_1(t),\dots,x_N(t)]},$ where $x_i(t)$ is the infection status of individual $i$ at time $t,$ with $x_i(t)=0$ and $x_i(t)=1$ representing susceptible and infected states, respectively.
The collection of state vectors at times $t=0,1,\dots,T$ is matrix $\boldX=[\boldx(0),\boldx(1),\dots,\boldx(T)].$

From previous work \cite{landry_reconstructing_2024}, the likelihood of a series of states $\boldX$ given network adjacency matrix $\boldA$ (assumed to be known) is
\begin{multline}
    P(\bm{X}\mid \bm{A},\gamma,\{\boldc_i\}) \propto \prod_{i=1}^N\prod_{\nu=1}^{N-1} c_i(\nu)^{M_{i, \nu}}(1-c_i(\nu))^{N_{i,\nu}},
\end{multline}
where $\gamma$ is the recovery rate,  $\boldc_i$ is the nonparametric contagion vector associated with node $i$, and $M_{i,\nu}$ and $N_{i,\nu}$ are the number of infection and non-infection events, respectively, for node $i$ when it has $\nu$ infected neighbors.
The associated matrices, $\bm{M}$ and $\bm{N}$, fully summarize the dynamics and are the only inputs required for inference.

We model node-specific contagion vectors $\boldc_i$ as draws from a finite mixture of nonparametric kernel modes.
That is, each node $i$ belongs to a single latent kernel class $z_i\in\{1,\dots,K\}$ and its logit-scale kernel $\tildec_i(\nu)=\operatorname{logit}c_i(\nu)\in\bbR$ varies around a class-level logit-scale mode $\tildemu_{z_i}(\nu)$,
$$
\tildec_i(\nu)\sim\mathrm{Normal}\,\left(\tildemu_{z_i}(\nu),\sigma\right).
$$

We constrain each $\mu_k$ to be monotonic in $\nu$, but our results hold without this assumption for sufficient data.
Details on this constraint and an expanded discussion of the mixture model are in Appendix A.

\paragraph{}\textit{Emergent complexity in networks with community structure}---We examine conditions under which emergent contagion complexity may arise by first considering networks generated from a two-block stochastic block model (SBM)~\cite{jerrum_metropolis_1998,karrer_stochastic_2011,landry_opinion_2023}.
This model has $N$ nodes split evenly between blocks, an expected mean degree $\langle k\rangle$, and within- and between-block edge probabilities of $p_{\mathrm{in}}=(1+\varepsilon)\langle k\rangle/(N-1),$ and $p_{\mathrm{out}}=(1-\varepsilon)\langle k\rangle/(N-1)$, respectively~\cite{landry_opinion_2023}.
In this model, $\varepsilon$ controls community separation; when $\epsilon=0$, we obtain an \ER network and when $\epsilon=1$, we obtain a network with two disconnected communities.
A network generated from this model with $N=256$, $\langle k\rangle=18$, and a separation of $\varepsilon=0.9$ is shown in \Cref{fig:illustration}(a).

Each node is assigned simple kernel
\begin{equation}\label{eq:simple_kernel}
c^{(g)}(\nu)=1-(1-\beta_g)^\nu,
\end{equation}
where $g\in\{1,2\}$ is the node's community label.
We let $\beta_1\leq\beta_2$ so that the only mechanistic difference between communities is their per-contact infectivity.
In \Cref{fig:illustration}, $\gamma=0.1$, $\beta_1=0.01$, and $\beta_2=0.04$.
When this community-level heterogeneity is ignored, and a global (\textit{i.e.}, single-component, $K=1$) nonparametric kernel mode is inferred, the estimate appears sigmoidal, as seen in \Cref{fig:illustration}(b), which is often associated with complex contagion~\cite{dodds_universal_2004,aiyappa_emergence_2024}.
In this case, however, both ground-truth kernels are simple.

\begin{figure}[bht]
    \centering
    \includegraphics[width=3.35in]{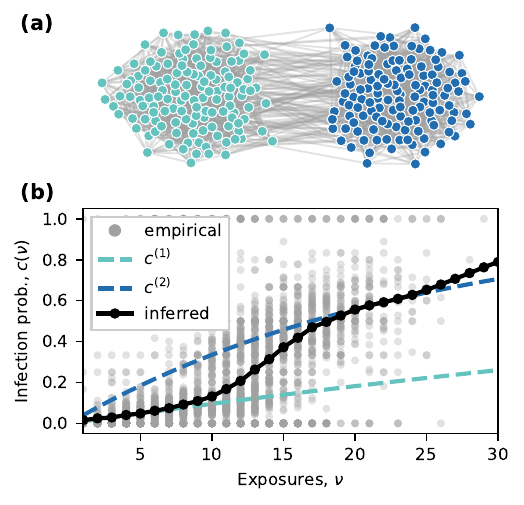}
    \caption{
        \textbf{Mixtures of simple contagions appear complex when heterogeneous infection probabilities are correlated.}
        (a) A two-community stochastic block model with $N=256$, $\langle k\rangle=18,$ and $\varepsilon=0.9$. Node colors indicate each individual's (unknown) underlying simple infection kernel---one of $c^{(1)}(\nu)$ or $c^{(2)}(\nu)$, shown in panel b.
        (b) Infection probability $c(\nu)$ given that $\nu$ neighbors are infected. The true underlying simple contagion kernels, $c^{(1)}(\nu)$ and $c^{(2)}(\nu)$, are of the form $1-(1-\beta_g)^{\nu}$ with $\beta_1=0.01$ and $\beta_2=0.04$. Inferring a global kernel mode yields the solid black curve, which exhibits a sigmoidal shape commonly associated with complex contagion kernels.
    }
    \label{fig:illustration}
\end{figure}

The apparent complexity of this global kernel arises from a correlation between the exposure level, $\nu$, and the latent community-specific infectivity.
Infections at low $\nu$-values come predominantly from $\beta_1$ nodes, which sustain fewer infections within their community, while infections at high $\nu$ are dominated by $\beta_2$ nodes, which sustain more infections within their community.
Therefore, fitting a single kernel mode to these data averages over the underlying simple kernels in an exposure-dependent way, and we recover a sigmoidal kernel.
This is a network-structured analogue of the nonlinear bias identified in Ref.~\cite{st-onge_nonlinear_2024}, where apparent complexity emerges because exposure becomes a proxy for unobserved transmission heterogeneity.

In our model, the correlation between individual kernels and community structure is tunable, allowing us to explore how aligned kernel and community assignments must be for the global kernel to exhibit emergent complexity.
We measure the complexity of an inferred contagion kernel by quantifying its departure from a simple contagion form [Eq.~\eqref{eq:simple_kernel}], adjusted to account for posterior uncertainty.
Specifically, we compute the Mahalanobis distance from a simple reference kernel parameterized by the posterior mean infectivity $\bar{\beta}_{\mathrm{sc}}$ of a simple model inferred from the same data.
We subtract the contribution expected from posterior spread alone and divide by that posterior-uncertainty baseline.
The resulting \textit{complexity score}, $D$, measures the excess squared distance from the simple contagion reference as a fraction of the posterior-uncertainty baseline.
That is, $D=0$ indicates no departure beyond posterior spread, while $D\geq 1$ indicates that the systematic departure is at least as large as the posterior-uncertainty baseline.
For more details, see Appendix B.

The complexity score allows us to find sufficient structural and dynamical conditions under which exposure becomes informative about latent infectivity in this two-block SBM. 
We first explore the influence of community separation and kernel-community alignment. 
Here, $\omega\in[0,0.5]$ represents the probability that a node's kernel assignment is "flipped" relative to its community label.
Kernels are perfectly aligned with communities when $\omega=0$ [as in \Cref{fig:illustration}(a)]; kernels are assigned independent of community labels when $\omega=0.5$.
\Cref{fig:sbm_sweeps}(a) shows that emergent complexity requires both ingredients.
When communities are weakly separated, or when kernels are assigned sufficiently at random across communities, the global kernel mode remains close to a simple contagion.
As $\epsilon$ increases and $\omega$ decreases, the exposure level, $\nu$, becomes increasingly predictive of the latent kernel, and the inferred global kernel no longer appears simple.

At fixed $\varepsilon=0.9$, the inferred kernel modes demonstrate this transition as $\omega$ decreases [\Cref{fig:sbm_sweeps}(b)].
When kernel assignments are randomized across communities ($\omega=0.5$), the global mode is close to a pointwise average of the two underlying kernels. 
As kernel assignments become increasingly aligned with the underlying communities, this average becomes exposure-dependent, forcing a transition between the $\beta_1$ and $\beta_2$ kernels;
the result is a sigmoidal kernel.

\begin{figure}[t!]
    \includegraphics[width=3.375in]{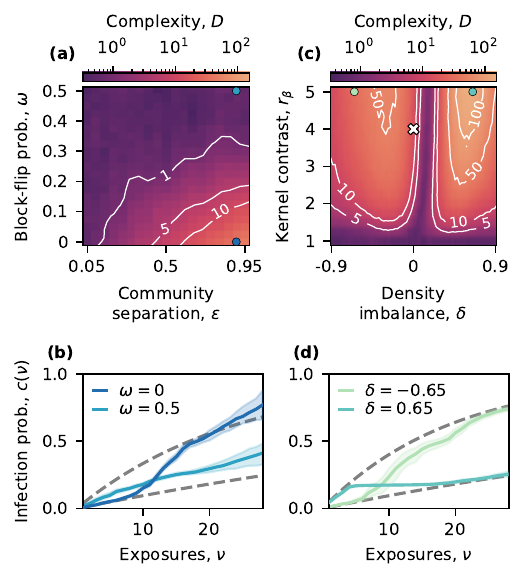}
    \caption{
        \textbf{Emergent complexity in the two-block $(\epsilon,\delta)$-stochastic block model with $N=256$, $\langle k\rangle=18$.} Each heatmap cell averages 10 simulations. (a) The complexity score, $D$, of inferred global kernel mode with respect to community separation ($\varepsilon$) and block-flip probability ($\omega$). We fix $\beta_1=0.01$ and $\beta_2=0.04.$ (b) The posterior mean global kernel modes at fixed $\varepsilon=0.9$ for varying $\omega$, with 90\% highest density interval (HDI) shading. Dashed curves show the ground-truth simple kernels. Kernels correspond to $(\varepsilon, \omega)$ combinations marked in panel (a). (c) The complexity score, $D$, of the inferred global kernel mode with respect to density imbalance $\delta$ and kernel contrast $r_\beta\equiv\beta_2/\beta_1$ with fixed $\varepsilon=0.9$ and $\omega=0.1$. The $(\delta,r_\beta)$ combination used in (a) is marked. (d) The posterior mean global kernel modes at fixed $r_\beta=5$ and two values of $\delta$, with 90\% HDI shading. Kernels correspond to $(r_{\beta}, \delta)$ combinations marked in panel (c). 
    }
    \label{fig:sbm_sweeps}
\end{figure}

This experiment assumes that the two communities have equal expected degrees, which is rarely the case in empirical networks \cite{leskovec_statistical_2008}.
To test whether the same mechanism persists when the communities have unequal densities, we introduce density imbalance parameter $\delta\in[-1, 1]$. 
Keeping the between-block probability $p_\mathrm{out}$ fixed, we multiply the within-block probability $p_\mathrm{in}$ by $(1+\delta)$ in block 1 and by $(1-\delta)$ in block 2 so that $\delta>0$ makes block 1 denser than block 2, and the opposite is true for $\delta<0$.
We fix $\varepsilon=0.9$ and $\omega=0.1$, values for which \Cref{fig:sbm_sweeps}(a) predicts emergent complexity, and vary both $\delta$ and the kernel contrast $r_\beta\equiv\beta_2/\beta_1.$
\Cref{fig:sbm_sweeps}(c) shows that---unsurprisingly---a sufficiently large contrast between the two simple kernels is necessary. 
When $r_\beta$ is close to 1, the population is effectively homogeneous, and the global kernel remains simple. 
For larger $r_\beta$, density imbalance can greatly amplify emergent complexity by separating the exposure distributions of the two latent kernel classes, making $\nu$ more informative about community-specific infectivity.

The two high-complexity regions in \Cref{fig:sbm_sweeps}(c) correspond to distinct non-simple kernel shapes.
For $\delta<0,$ the density imbalance reinforces the mechanism described above: high-exposure observations are concentrated among $\beta_2$ nodes, producing the same sigmoidal aggregate kernel [\Cref{fig:sbm_sweeps}(d)].
For $\delta>0,$ the $\beta_1$ community is denser than the $\beta_2$ community.
Consequently, the inferred kernel grows quickly at low exposure, flattens, and rises again---more slowly---at larger $\nu$.
Between these two regimes, we see a low-complexity region whose location shifts to larger $\delta$ as $r_\beta$ increases.
Increasing $\beta_2$ increases the typical exposure level in the $\beta_2$ block, so a larger (positive) $\delta$ is needed to densify the $\beta_1$ block and restore overlap between the two exposure distributions. 
See the Supplemental Material for more details.

Notably, emergent complexity is not limited to the canonical, threshold-like signal of peer reinforcement [\Cref{fig:sbm_sweeps}(d)]: correlated heterogeneity generates a broader class of global kernels whose shape depends on how network structure maps latent contagion types onto exposure levels.
The same mechanism also arises beyond idealized block models: in the Supplemental Material, we observe emergent complexity on an empirical network.

\paragraph{}\textit{Disentangling mechanistic and emergent complexity}---The complexity score identifies when an inferred kernel mode departs from the simple contagion family, but it does not distinguish emergent complexity from mechanistic complexity.
A non-simple global kernel mode could reflect a genuinely complex contagion rule shared by all nodes, or it could arise from aggregating over heterogeneous subpopulations with simpler rules.
To distinguish these possibilities, we use node-level cross-validation (CV) to select the number of mixture components $K$.
\begin{figure}
    \centering
    \includegraphics[width=3.375in]{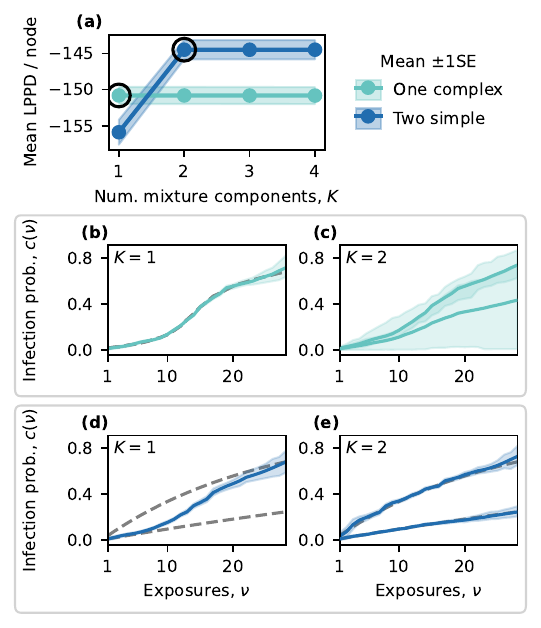}
    \caption{\textbf{Cross-validation distinguishes a single complex kernel from a heterogeneous mixture of simple kernels in a two-block SBM with $\varepsilon=0.9$.} (a) Mean held-out log pointwise predictive density (LPPD) per node under 5-fold cross-validation with respect to number of mixture components, $K$. For data generated by one complex kernel, the 1-SE rule selects $K=1$; for data generated by two simple kernels, the 1-SE rule selects $K=2$. (b,c) Inferred kernel mode(s) with 90\% HDI shading for data generated by one complex kernel, shown in gray. (d,e) Inferred kernel mode(s) with 90\% HDI shading for data generated by two simple kernels, shown in gray.
    }
    \label{fig:cv_LPPD}
\end{figure}

For each candidate $K=1,\dots,K_{\max}$, we perform $k$-fold cross-validation over nodes.
We compute the exposure-count matrices $\bm{M}$ and $\bm{N}$ from the full observed time series on the complete network, then hold out rows of these matrices from the likelihood.
Thus, held-out nodes' transition counts are excluded from training, but their observed states still contribute to neighboring nodes' exposure counts. 
We score each model by the mean held-out log pointwise predictive density \cite{vehtari_practical_2017} per node across folds, marginalizing over each held-out node's latent component assignment $z_i$.
Because unsupported components typically receive negligible weight or duplicate existing modes, CV scores tend to plateau.
We therefore select the smallest $K$ whose CV score lies within one standard error of the maximum \cite{hastie_model_2009}; see Appendix C for more information.

\Cref{fig:cv_LPPD} demonstrates that node-level CV can distinguish a shared non-simple kernel from a heterogeneous mixture of simpler kernels in two-block SBMs with strong community separation $(\varepsilon=0.9)$ and strongly aligned kernel assignments $(\omega=0.1)$.
When the data are generated by a single complex kernel, additional mixture components do not improve held-out predictive performance, and the 1-SE rule consistently selects $K=1$ [\Cref{fig:cv_LPPD}(a)].
The selected model therefore recovers a single population-level non-simple contagion rule [\Cref{fig:cv_LPPD}(b)].
In the over-specified $K=2$ model, one component recovers the same kernel, while the extra component is only weakly identified: its posterior is diffuse, and its mixture weight is negligible [\Cref{fig:cv_LPPD}(c)].

In contrast, when the data are generated by two simple kernels, the CV score increases sharply from $K=1$ to $K=2$ and then plateaus [\Cref{fig:cv_LPPD}(a)].
The $K=1$ model aggregates the two subpopulations into an exposure-dependent average, producing a complex global kernel [\Cref{fig:cv_LPPD}(d)].
The $K=2$ model instead recovers the underlying simple component modes [\Cref{fig:cv_LPPD}(e)].
These simulations demonstrate that cross-validation can disentangle mechanistic and emergent complexity: mechanistic complexity is supported when a single non-simple kernel is sufficient for held-out prediction, whereas emergent complexity is supported when multiple simpler components are required.

\paragraph{}\textit{Discussion}---In this Letter, we have demonstrated that inferential frameworks can recover contagion rules that suggest peer reinforcement is at play---the classical "complex contagion" paradigm---when, in reality, the underlying mechanism is heterogeneous simple contagion.
This suggests that the current understanding of complex contagion should be expanded to encompass not only mechanistic complexity, often characterized by sigmoidal contagion kernels, but also emergent complexity, characterized by correlated heterogeneity.
This work introduces a \textit{complexity score}, a measure of complexity quantifying the amount by which a contagion kernel deviates from the simple contagion family.
We use this measure to explore network structures and dynamics that exhibit emergent complexity.
This measure, however, cannot capture the full spectrum of complexity; as illustrated in \Cref{fig:sbm_sweeps}(d), two qualitatively different kernels with large complexity scores arise.
Emergent complexity arises when community structure and kernel assignments are sufficiently aligned; density differences between communities can then amplify the effect or reshape the inferred global kernel.
Finally, we have shown that it is possible to distinguish mechanistic from emergent complexity using a Bayesian inferential approach coupled with cross-validation.

While we have demonstrated that emergent complexity can, in principle, occur, here we have focused on the simplest case: two communities and two simple contagion kernels.
These results raise several questions about the generality, detectability, and empirical relevance of emergent complexity.
First, we must extend our framework to an arbitrary number of communities and kernels.
What does the resulting kernel look like?
Are we still able to disentangle the underlying kernels?
Second, while we have suggested that emergent complexity may plausibly explain contagions that look complex, this phenomenon has not yet been identified in systems with empirical dynamics.
Nonetheless, this study is a meaningful step forward in measuring and inferring complex and heterogeneous contagion kernels.

\paragraph{}\textit{Data availability}---The data and code that support the findings of this study are archived on Zenodo at \url{https://doi.org/10.5281/zenodo.21632561}. Any current unreleased code is available on GitHub at \url{https://github.com/kaiser-dan/heterogeneous-infection-kernels}.

\acknowledgments{
    The authors acknowledge support from the National Institutes of Health (1P20 GM125498-01 Centers of Biomedical Research Excellence Award, J.-G.Y., N.W.L., \& L.H.-D.), from The National Science Foundation (award \#2419733, J.-G.Y. \& L.H.-D.) and from the University of Virginia Prominence-to-Preeminence (P2PE) STEM Targeted Initiatives Fund, SIF176A Contagion Science (N.W.L. and D.K.). 
    The authors acknowledge Research Computing (rc.virginia.edu) at The University of Virginia for providing computational resources and technical support that contributed to the results reported in this publication.
    The authors would like to thank Sichen Jin and Clio Andris for providing data.
    }

\bibliography{references}

\newpage

\onecolumngrid
\begin{center}
\textbf{End Matter}\\[0.75cm]
\end{center}
\twocolumngrid

\textit{Appendix A: Mixture model.} To pool information across nodes while accommodating population heterogeneity, we model node $i$'s contagion kernel $\boldc_i$ as drawn from a finite mixture of $K$ latent components.
To allow for more flexible modeling, we work throughout on the logit scale, defining
$$\Tilde{c}_i(\nu)\equiv\mathrm{logit}\ c_i(\nu)\in\mathbb{R}.$$
Node $i$ is assigned to component $z_i\in\{1,\dots,K\}$ with probability $\pi_k=P(z_i=k),$ where $\bm{\pi}\sim\mathrm{Dirichlet}(\bm{1}_K).$
Nodes assigned to the same mixture component share a common \textit{kernel mode} $\bm{\mu}_k$---a subpopulation-level contagion kernel reflecting characteristic behavior for that group---around which individual node kernels vary on the logit scale:
$$
\tildec_i(\nu)\sim\mathrm{Normal}\,\left(\tildemu_{z_i}(\nu),\sigma\right),
$$
where $\sigma$ is a global scale parameter controlling the degree of node-level heterogeneity within each component.

We construct each component $k$'s kernel mode to be non-decreasing in $\nu,$ reflecting our assumption that additional infectious contacts cannot reduce infection risk.
Specifically, we form the cumulative sum of a Dirichlet-distributed increment vector $\bm{\Delta}_k\sim\mathrm{Dirichlet}(\kappa\bm{1}_{k_{\max}})$ scaled by a component-specific ceiling probability $p_k\in[0,1]:$
$$\mu_k(\nu)=p_k\sum_{j=1}^\nu\Delta_{k,j}$$

The concentration parameter $\kappa$---fixed at 1.0 throughout our experiments---controls the smoothness of kernel modes: small values of $\kappa$ allow highly non-uniform kernels where infection probability increases rapidly over a narrow range of $\nu,$ while large values favor smoother, more gradual kernels.
Note that we truncate the kernel domain at $k_{\max}\leq N-1$ for simplicity.

The ceiling $p_k$ directly controls the maximum infection probability attained under component $k,$ encoding the assumption that infection risk saturates below certainty even at high exposure.
We place a normal prior on the logit-scale ceiling $\ell_k\equiv\mathrm{logit}\ p_k$ centered at the empirical log-odds of infection in the data. 
This places prior mass in a region consistent with the scale of the data and prevents the prior from dominating the likelihood when infection rates are close to zero or one.

\textit{Appendix B: Complexity score.} We quantify the apparent complexity of an inferred contagion kernel mode by measuring its departure---adjusted for posterior uncertainty---from a reference kernel.
In this work, the reference kernel is a simple contagion kernel, but the same construction can be applied to any hypothesized kernel shape.

Let $\Tilde{\boldmu}^{(s)}\in\mathbb{R}^{k_{\max}}$ denote posterior draw $s$, $s=1,\dots,S$, of the logit-scale kernel mode, and let $\Tilde{\boldmu}_{\mathrm{sc}}$ denote the logit-scale simple contagion reference,
$$\Tilde{\boldmu}_{\mathrm{sc}}(\nu)=\mathrm{logit}\,\left[1-(1-\bar{\beta}_{\mathrm{sc}})^\nu\right],$$
where $\bar{\beta}_{\mathrm{sc}}$ is the posterior mean infectivity under the simple contagion model fit to the same data.
All distances are computed on the logit scale.

For each posterior draw, define the regularized Mahalanobis distance
$$
Q^{(s)}\equiv 
\trans{\left(\Tilde{\bm{\mu}}^{(s)}-\Tilde{\bm{\mu}}_{\text{sc}}\right)}
\bm{P}
\left(\Tilde{\bm{\mu}}^{(s)}-\Tilde{\bm{\mu}}_{\text{sc}}\right),
$$
where
$$\bm{P}=\left(\Hat{\bm{\Sigma}}+\lambda\frac{\mathrm{tr}(\Hat{\bm{\Sigma}})}{k_{\max}}\bm{I}\right)^{-1}.$$
Here, $\Hat{\bm{\Sigma}}$ is the posterior sample covariance of $\Tilde{\boldmu}$ and $\bm{P}$ is a regularized inverse covariance.
We set $\lambda\equiv\alpha_{\mathrm{LW}}/(1-\alpha_{\mathrm{LW}}),$ where $\alpha_{\mathrm{LW}}$ is the Ledoit-Wolf shrinkage estimate \cite{ledoit_well_2004}, to preserve the posterior covariance geometry while preventing poorly estimated near-zero covariance eigenvalues from receiving unbounded weight.

To interpret $Q^{(s)}$, write a posterior draw as
$$
\Tilde{\boldmu}=\bar{\boldmu}+\bm{\epsilon},
\quad \bar{\boldmu}=\mathbb{E}[\Tilde{\boldmu}],
\quad \mathbb{E}[\bm{\epsilon}]=\bm{0},
\quad \mathrm{Cov}(\bm{\epsilon})=\bm{\Sigma}.
$$ 
Then
\begin{align*}
    \mathbb{E}\left[\trans{\left(\Tilde{\bm{\mu}}-\Tilde{\bm{\mu}}_{\text{sc}}\right)}\bm{P}\left(\Tilde{\bm{\mu}}-\Tilde{\bm{\mu}}_{\text{sc}}\right)\right]
    =&\;
    \trans{\left(\bar{\bm{\mu}}-\Tilde{\bm{\mu}}_{\text{sc}}\right)}\bm{P}\left(\bar{\bm{\mu}}-\Tilde{\bm{\mu}}_{\text{sc}}\right)\\
    &+\mathrm{tr}(\bm{P}\bm{\Sigma}).
\end{align*}
The first term measures displacement of the posterior mean from the reference, while the trace term is the distance expected from posterior spread alone under the regularized Mahalanobis metric.
We therefore define our complexity score as
$$
D\equiv\frac{\langle Q\rangle_{s}-\mathrm{tr}(\bm{P}\hat{\bm{\Sigma}})}{\mathrm{tr}(\bm{P}\hat{\bm{\Sigma}})},\qquad \langle Q\rangle_{s}=S^{-1}\sum_{s=1}^S Q^{(s)}.
$$
Thus, $D$ is a dimensionless signal-to-uncertainty ratio on the squared Mahalanobis scale.
The value $D=0$ indicates no systematic departure from the reference kernel beyond posterior spread, while $D=1$ indicates that the excess (squared) departure is equal to the uncertainty baseline.
Equivalently, the posterior mean squared Mahalanobis distance from the reference is $1+D$ times the distance expected from posterior uncertainty alone. 

As a limiting case, suppose the logit-scale kernel mode $\Tilde{\boldmu}$ were known exactly, i.e., $\bm{\Sigma}=\bm{0}.$
Apparent complexity relative to the simple contagion reference could be measured in a straightforward manner as the mean squared logit-scale deviation:
$$D_0=\frac{1}{k_{\max}}\trans{\left(\Tilde{\bm{\mu}}-\Tilde{\bm{\mu}}_{\text{sc}}\right)}
\left(\Tilde{\bm{\mu}}-\Tilde{\bm{\mu}}_{\text{sc}}\right)$$
The score $D$ is the uncertainty-adjusted analogue: it replaces the identity geometry with a posterior-uncertainty-aware geometry, subtracts the distance expected from posterior spread alone, and normalizes the remaining systematic departure by that uncertainty baseline.

Finally, $D$ can be calibrated by posterior predictive simulation under a fitted simple contagion null, as described in the Supplemental Material.
This calibration is useful because even a truly simple contagion can yield a positive estimated complexity score when the time series is short and the kernel is inferred rather than observed.

\textit{Appendix C: Node-level cross-validation.} We use node-level cross-validation (CV) to select the number of mixture components, $K$.
For each fold, the exposure-count matrices $\bm{M}$ and $\bm{N}$ are formed once on the full network, and held-out nodes are excluded only by dropping their rows from the training likelihood. Thus, held-out nodes are not removed from the dynamical process: their observed states still contribute to the exposure counts of the training nodes.

Recall that $M_{i,\nu}$ and $N_{i,\nu}$ count, respectively, the infection and non-infection events at node $i$ while it had $\nu$ infected neighbors.
After fitting on the training set, we score the held-out fold by its log pointwise predictive density (LPPD) \cite{vehtari_practical_2017}:
$$
\text{LPPD}_j = \sum_{i \in \mathcal{F}_j} \log
\hat{p}(M_i, N_i \mid\{M_\ell, N_\ell: \ell\in\mathcal{F}_{-j}\}),
$$
where
\begin{align*}
\hat p(M_i,N_i)
=&\; \frac{1}{S}\sum_{s=1}^{S}\sum_{k=1}^{K}\bigg{[}\pi_k^{(s)}\\
&\times\prod_{\nu=1}^{k_{\max}} \mu_k^{(s)}(\nu)^{M_{i,\nu}}
\bigl(1-\mu_k^{(s)}(\nu)\bigr)^{N_{i,\nu}}\bigg{]},
\end{align*}
with $\{\bm{\pi}^{(s)},\bm{\mu}^{(s)}\}_{s=1}^S$ drawn from the posterior given
$\{M_\ell, N_\ell: \ell\in\mathcal{F}_{-j}\}.$
The predictive therefore depends only on the population-level kernels $\mu_k$ and weights $\pi_k$; no per-node latent quantity enters.

The overall CV score is the mean LPPD per held-out node.
We repeat this procedure for $K=1,2,\dots,K_{\max}$ mixture components and apply the one-standard-error (1-SE) rule: we select the smallest $K$ whose mean LPPD lies within one standard error of the maximum \cite{hastie_model_2009}.
This parsimony criterion is useful here because unsupported extra components cannot be tuned to held-out node-specific fluctuations; the held-out predictive depends only on subpopulation-level kernel modes and mixture weights.
In practice, extra components tend to receive negligible weight or duplicate existing components, so the CV curve plateaus---with at most a slight downward drift from the extra population-level parameters---once $K$ captures the population heterogeneity.
In our simulations, these plateaus begin at the true $K$, and the 1-SE rule recovers the true number of components across replicates; see the Supplemental Material.

\clearpage
\title{Supplemental Material for \textit{Emergent contagion complexity: Disentangling mechanistic complexity from correlated heterogeneity}}

\setlength{\parskip}{0pt}
\maketitle
\onecolumngrid

\renewcommand{\thefigure}{S\arabic{figure}}
\setcounter{figure}{0}
\renewcommand{\thetable}{S\arabic{table}}
\setcounter{table}{0}
\renewcommand{\theequation}{S\arabic{equation}}
\setcounter{equation}{0}

\setlength{\parskip}{4pt}

\textbf{Cross validation for selecting the number of mixture components.} As described in Appendix C, we use five-fold node-level cross-validation (CV) and the one-standard-error rule to select the number of mixture components $K$.
Here, we show that this procedure is stable across simulation replicates (20 for each ground truth data generating process).
Throughout, we consider networks drawn from our $(\epsilon,\delta)$-stochastic block model with $N=256,$ $\langle k\rangle$=18, $\varepsilon=0.9,$ and $\delta=0.0.$

\begin{figure}[thbp]
    \centering
    \includegraphics[width=6.5in]{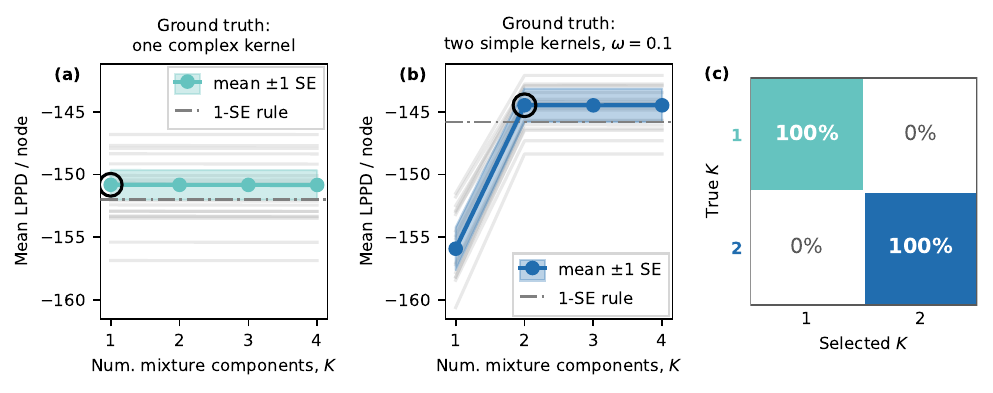}
    \caption{\textbf{Cross-validation recovers the true number of kernel components.} (a) Mean held-out LPPD per node versus number of mixture components, $K$, for data generated by one complex kernel. (b) Corresponding CV curves for data generated by two simple kernels with block-flip probability $\omega=0.1$. Gray curves show individual simulation replicates; colored points and error bars show the mean $\pm 1$ SE for one highlighted simulation. Dashed horizontal lines indicate the 1-SE selection threshold. (c) Confusion matrix comparing the true and selected number of components.
}
    \label{fig:supplementary:cv}
\end{figure}

For data generated by a single, complex kernel---the sigmoidal kernel recovered in \Cref{fig:illustration}---the CV score curves consistently plateau at $K=1$, indicating that additional components do not improve predictive performance [\Cref{fig:supplementary:cv}(a)].
For data generated by a mixture of two simple kernels with $\beta_1=0.01$ and $\beta_2=0.04$---assigned to communities 1 and 2, respectively, with block-flip probability $\omega=0.1$---the CV score curves consistently increase from $K=1$ to $K=2$ and then plateau [\Cref{fig:supplementary:cv}(b)].
Applying the one-standard-error rule recovers the true number of kernel components across all simulations [\Cref{fig:supplementary:cv}(c)].

\textbf{Empirical results.} To complement the experiments on synthetic contact structures presented in \Cref{fig:sbm_sweeps}, we next show that emergent complexity can arise with synthetic SIS dynamics on an empirical network.
We use a network of United States representatives in the 118th Congress, connecting two representatives if they agreed on at least 980 roll-call votes.
The data is gathered from publicly available records \cite{congress}.
The resulting network is undirected, unweighted, and without multi-edges or self-loops.
We restrict our analysis to the 107 representatives in the largest connected component of this network.
We find two communities of representatives using a spectral method, bisecting the network according to the sign of the entries of the second-largest eigenvalue of the modularity matrix \cite{newman_modularity_2006}.
Nodes in the denser of the two communities are assigned with probability $1-\omega$ an underlying simple kernel with $\beta_2 = 0.05$; nodes in the sparser community are similarly assigned a simple kernel with $\beta_1 = 0.01$. 
We assume a constant healing rate of $\gamma=0.1$.

\begin{figure}[thb]
    \centering
    \includegraphics[width=6.5in]{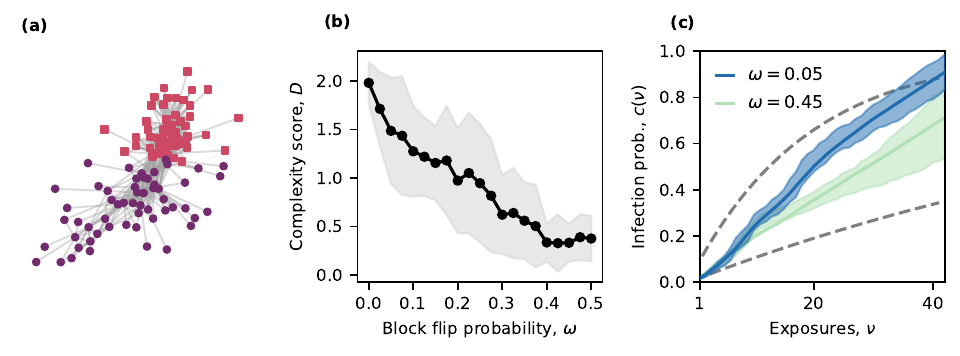}
    \caption{
        \textbf{Emergent complexity on an empirical network.} 
        (a) The congressional network with node color and shape indicating its community label.
        (b) The complexity score, $D$, with respect to the block-flip probability.
        We plot the average of 10 repetitions and $\pm$ 1 standard deviation.
        (c) Representative samples from the posterior kernel with underlying ground-truth kernels plotted in gray.
    }
    \label{fig:supplementary:congress}
\end{figure}

With sufficiently small $\omega$---that is, sufficiently high kernel-community alignment---the inferred global kernel has significant complexity [\Cref{fig:supplementary:congress}, \Cref{tab:supplementary:ppc}].
We assess significance with a posterior predictive null for global simple contagion with shared .
We construct a null distribution by posterior predictive simulation: for each replicate $r=1,\dots,R,$ we draw $\beta^{(r)}$ from the posterior of the simple contagion model fit to the observed data and simulate a new dataset on the same network with the same observation window and recovery probability $\gamma$ used in the observed analysis.
We then fit the nonparametric kernel model to the replicate dataset and compute the corresponding complexity score $D_{\mathrm{rep}}^{(r)}$ with the simple contagion reference kernel parameterized by $\beta^{(r)}.$
The dimensionless ridge parameter $\lambda$ used in computing $D_{\mathrm{rep}}^{(r)}$ is chosen once---from the observed data---and then held fixed across replicate analyses.

The posterior predictive $p$-value is then
$$p_{\mathrm{ppc}}=\frac{1+\sum_{r=1}^R\mathbbm{1}\left[D_{\mathrm{rep}}^{(r)}\geq D_{\mathrm{obs}}\right]}{R+1},$$
the fraction of null replicates whose complexity is at least as large as that observed in the original dataset (with a finite-replicate correction).
Such calibration is necessary because finite time series, stochastic simulation noise, posterior uncertainty, and covariance estimation can all produce nonzero complexity even when the data-generating kernel is simple.
Thus, $p_{\mathrm{ppc}}$ measures whether the observed departure from the simple contagion family is larger than expected under a fitted simple contagion null model.

The observed complexity score is significant for small $\omega$ ($p<0.05$; \Cref{tab:supplementary:ppc}), indicating more apparent complexity than expected under a homogeneous simple contagion process on the same network.

\begin{table}
\centering
\begin{tabular}{p{2cm}p{5cm}}
\toprule
$\omega$ & Posterior predictive $p$-value \\
\midrule
0.0 & $0.0198^{*}$ \\
0.1 & $0.0099^{**}$ \\
0.2 & $0.0099^{**}$ \\
0.3 & 1.0000 \\
0.4 & 0.9208 \\
0.5 & 0.9909 \\
\bottomrule
\end{tabular}
\caption{\label{tab:supplementary:ppc}Posterior predictive $p$-values for kernels inferred from the Congress agreement dataset. $^{*}p \le 0.05$, $^{**}p \le 0.01$}
\end{table}

This result is consistent with the SBM results above but shows that emergent complexity is not specific to idealized block models: the same exposure-dependent aggregation can arise on an empirical network with naturally occurring community structure.
Moreover, the inferred global kernel is not strongly sigmoidal, reinforcing that emergent complexity can produce nonstandard departures from the simple contagion family.

\textbf{The basic reproduction number for two simple kernels and the $(\epsilon, \delta)$-stochastic block model.}
To derive the basic reproduction number for the model used in generating the results in Fig.~\ref{fig:sbm_sweeps}, we employ a mean-field approach.
For a network of size $N$ and a fixed mean degree $\langle k\rangle$, the $(\epsilon,\delta)$-stochastic block model (SBM) is a 2-parameter model, where the community separation parameter, $\epsilon$, controls the number of links between the two blocks~\cite{jerrum_metropolis_1998}, each of size $N/2$, and the density imbalance parameter, $\delta$, controls the relative densities between the two blocks.
When $\delta=0$, we recover the standard planted partition model; in this case, when $\epsilon=0$, we recover an \ER network.
When $\epsilon=1$, we obtain two isolated communities.
When $\delta > 0$, block 1 is denser than block 2, and when $\delta < 0$, the opposite is true.
Conditioned on node labels, between-block edges occur with probability $p_{12}=p_{21}=(1-\varepsilon)\langle k\rangle/(N-1)$, while within-block edges occur with probabilities $p_{11}=(1+\delta)(1+\varepsilon)\langle k \rangle/(N-1)$ and $p_{22}=(1-\delta)(1+\varepsilon)\langle k \rangle/(N-1)$ in blocks 1 and 2, respectively.

We define $x_1$ and $x_2$ as the fraction of infected nodes in communities 1 and 2, respectively, and, following the same steps as in Ref.~\cite{landry_opinion_2023}, we obtain the following system of equations:
\begin{align}
\frac{dx_1}{dt}&=-\gamma x_1 + \bar{\beta}_1\frac{\langle k\rangle}{2} (1-x_1) \left[(1+\epsilon)(1+\delta)x_1 +  (1-\epsilon) x_2 \right],\\
\frac{dx_2}{dt}&=-\gamma x_2 + \bar{\beta}_2\frac{\langle k\rangle}{2} (1-x_2) \left[(1-\epsilon) x_1 + (1+\epsilon)(1-\delta)x_2 \right],
\end{align}
where $\bar{\beta}_1=(1-\omega)\beta_1 + \omega\beta_2$ and $\bar{\beta}_2=\omega\beta_1  + (1 - \omega) \beta_2$ denote the average infectivity in blocks 1 and 2, respectively.

Linearizing this system about the $\mathbf{0}$ equilibrium, we obtain the linear equation for the perturbations about this equilibrium,
\begin{align*}
\frac{d}{dt}
\begin{bmatrix}
x_1\\
x_2
\end{bmatrix}
&=
\left( -\gamma I + \frac{\langle k\rangle}{2}
\begin{bmatrix}
\bar{\beta}_1(1  + \epsilon)(1+\delta) & \bar{\beta}_1 (1-\epsilon)\\
\bar{\beta}_2 (1 - \epsilon) & \bar{\beta}_2 (1+\epsilon)(1-\delta)
\end{bmatrix}
\right)
\begin{bmatrix}
x_1\\
x_2
\end{bmatrix}.
\end{align*}
Then, letting $B$ denote the infection matrix above,
\begin{align*}
\text{tr}(B)=&\,\left(\frac{\langle k \rangle}{2} (1 + \epsilon) (1+\delta) \bar{\beta}_1\right) + \left(\frac{\langle k \rangle}{2} (1 + \epsilon) (1-\delta) \bar{\beta}_2\right)\\
=& \frac{\langle k\rangle}{2} (1+\epsilon)\left[\left(\bar{\beta}_1 + \bar{\beta}_2\right) + \delta\left(\bar{\beta}_1 - \bar{\beta}_2\right)\right]\\
\det(B) =&\, \left(\frac{\langle k \rangle}{2} (1 + \epsilon) (1+\delta) \bar{\beta}_1\right)\left(\frac{\langle k \rangle}{2} (1 + \epsilon) (1-\delta) \bar{\beta}_2\right)\\
&-\left(\frac{\langle k \rangle}{2} (1 - \epsilon)\bar{\beta}_1\right)\left(\frac{\langle k \rangle}{2} (1 - \epsilon)\bar{\beta}_2\right)\\
=&\,\frac{\langle k \rangle^2}{4} \bar{\beta}_1 \bar{\beta}_2\left[(1+\epsilon)^2(1-\delta^2) - (1 - \epsilon)^2\right].
\end{align*}
Then the eigenvalues of $-\gamma I + B$ are
\begin{align*}
\lambda =& -\gamma + \frac{\langle k\rangle}{4}(1+\epsilon)\left[\left(\bar{\beta}_1 + \bar{\beta}_2\right) + \delta\left(\bar{\beta}_1 - \bar{\beta}_2\right)\right]\\
&\pm \frac{\langle k \rangle}{4}\sqrt{(1+\epsilon)^2\left[\left(\bar{\beta}_1 - \bar{\beta}_2\right) + \delta\left(\bar{\beta}_1 + \bar{\beta}_2\right)\right]^2 + 4(1 - \epsilon)^2\bar{\beta}_1 \bar{\beta}_2},\\
=&\,-\gamma + \frac{\langle k\rangle}{4}(1+\epsilon)\Bigg{[}(\beta_1 + \beta_2) + \delta(1-2\omega)(\beta_1 - \beta_2)\\
&\pm\sqrt{ [(1 - 2\omega)(\beta_1 - \beta_2) + \delta(\beta_1 + \beta_2)]^2 + \left(\displaystyle\frac{1 - \epsilon}{1+\epsilon}\right)^2[(\beta_1 + \beta_2)^2 - (1 - 2\omega)^2(\beta_1-\beta_2)^2]}\;\Bigg{]},\\
\end{align*}
and the reproduction number is
\begin{align}\label{eq:r0}
R_0 =& \frac{\langle k \rangle(1+ \epsilon)}{4\gamma}\bigg{[}(\beta_1 + \beta_2) + \delta(1-2\omega)(\beta_1 - \beta_2)\nonumber \\
&+ \sqrt{ [(1 - 2\omega)(\beta_1 - \beta_2) + \delta(\beta_1 + \beta_2)]^2 + \left(\displaystyle\frac{1 - \epsilon}{1+\epsilon}\right)^2[(\beta_1 + \beta_2)^2 - (1 - 2\omega)^2(\beta_1-\beta_2)^2]}\bigg{]}.
\end{align}

\begin{figure}
    \centering
    \includegraphics[width=3.8in]{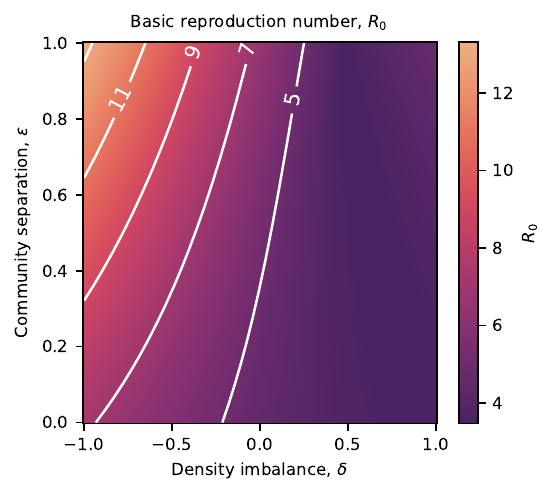}
    \caption{Analytical estimate of $R_0$ from \Cref{eq:r0} derived from our mean-field model. Here, as in \Cref{fig:sbm_sweeps}, $\langle k\rangle=18$. Additional parameter values are $\beta_1 \equiv 0.01, \beta_2 \equiv 0.04$, and $\omega\equiv 0.1$.}\label{fig:supplementary:reproduction_number}
\end{figure}

As seen in \Cref{fig:supplementary:reproduction_number}, the reproduction number is strongly dependent on the relative densities of the two communities.
Notice that negative $\delta$ values correspond to a large increase in the reproduction number.
This matches what we would expect from \Cref{fig:sbm_sweeps}(d), where the global inferred kernel follows the more infectious contagion kernel for small values of $\nu$.
For positive $\delta$ values, we see a decrease in the reproduction number, and this corresponds to the sigmoidal "complex" kernel in \Cref{fig:sbm_sweeps}(d).
As found in Ref.~\cite{gupta_interplay_2026}, when $\omega \neq 1/2$ and $\beta_1 \neq \beta_2$, the reproduction number depends on the strength of community structure.

\textbf{Mean-field analysis of the low-complexity region in the $(\varepsilon,\delta)$-stochastic block model.}
\Cref{fig:sbm_sweeps}(c) shows a low-complexity region whose location shifts to larger $\delta$ as the kernel contrast, $r_\beta$, increases.
Here we derive a mean-field predictor for the location of this low-complexity region based on the observation that a global kernel appears complex when exposure level $\nu$ is informative about a node's latent kernel class.

Let $L_{i,\nu}=M_{i,\nu}+N_{i,\nu}$ denote the number of exposure observations for a susceptible node $i$ at exposure level $\nu$.
An empirical estimate of the infection probability at exposure $\nu$ is
\begin{equation*}
    \hat{c}(\nu)=\frac{\sum_{i=1}^N M_{i,\nu}}{\sum_{i=1}^N L_{i,\nu}}.
\end{equation*}
For each kernel class $\ell\in\{1,2\}$, let $M_\nu^{(\ell)}$ and $L_\nu^{(\ell)}$ denote the number of infections and total susceptible exposure observations, respectively, among $\beta_\ell$ nodes at exposure level $\nu$.
We can rewrite $\hat{c}(\nu)$ as
$$
\hat{c}(\nu) 
= \frac{M_\nu^{(1)}}{L_\nu^{(1)}+L_\nu^{(2)}} + \frac{M_\nu^{(2)}}{L_\nu^{(1)}+L_\nu^{(2)}}\\
= [1-W_{\beta_2}(\nu)]\left(\frac{M_\nu^{(1)}}{L_\nu^{(1)}}\right)+W_{\beta_2}(\nu)\left(\frac{M_\nu^{(2)}}{L_\nu^{(2)}}\right)
$$
where 
\begin{equation}\label{eq:W2}
    W_{\beta_2}(\nu)=\frac{L_\nu^{(2)}}{L_\nu^{(1)} + L_\nu^{(2)}}
\end{equation}
is the fraction of susceptible observations at exposure level $\nu$ contributed by nodes with the high-infectivity ($\beta_2$) kernel.

From the form of $\hat{c}(\nu),$ we see that apparent complexity is controlled not only by the kernel contrast between $\beta_1$ and $\beta_2$ but also by how $W_{\beta_2}(\nu)$ varies over the observed exposure range.
If $W_{\beta_2}(\nu)$ is approximately constant in $\nu,$ the global kernel is close to a fixed convex combination of the two simple kernels and remains comparatively simple.
If $W_{\beta_2}(\nu)$ varies strongly with $\nu,$ the global estimate averages the two simple kernels in an exposure-dependent way, producing an apparent departure from the simple contagion family.

We use a block mean-field approximation to estimate $W_{\beta_2}(\nu)$ under the $(\varepsilon,\delta)$-SBM with network size $N$ and fixed mean degree $\langle k\rangle.$
Let 
$$\rho_{g,\ell}(t)=P(\text{a node in block $g$ with kernel $\ell$ is infected at time $t$}),$$
where $g,\ell\in\{1,2\}.$
The exposure distribution of a susceptible node depends on its structural block through the aggregate block prevalences:
\begin{equation}\label{eq:agg_block_prevalence}
    \bar{\rho}_1(t)=(1-\omega)\rho_{1,1}(t)+\omega\rho_{1,2}(t),\qquad\bar{\rho}_2(t)=\omega\rho_{2,1}(t)+(1-\omega)\rho_{2,2}(t),
\end{equation}
where, recall, a node in block 1 is assigned kernel 1 with probability $1-\omega$ and kernel 2 with probability $\omega$ and vice versa for a node in block 2.
The number of infected neighbors of a susceptible node in blocks 1 and 2, respectively, is approximated as Poisson with mean
\begin{equation}\label{eq:poisson_mean}
    \lambda_1(t)
    =
    \left(\frac{N}{2}-1\right)p_{11}\bar{\rho}_1(t)
    +
    \frac{N}{2}p_{12}\bar{\rho}_2(t),
    \qquad
    \lambda_2(t)
    =
    \frac{N}{2}p_{12}\bar{\rho}_1(t)
    +
    \left(\frac{N}{2}-1\right)p_{22}\bar{\rho}_2(t),
\end{equation}
where $p_{11}$ and $p_{22}$ are the within-block edge probabilities and $p_{12}$ is the between-block edge probability defined in the previous section.
The first term in $\lambda_1$ is the expected number of infected neighbors from within block 1 and the second term is the expected number of infected neighbors from block 2; the terms in $\lambda_2$ have a similar interpretation.

The mean infection probability for a susceptible node in block $g$ with simple kernel $c^{(\ell)}(\nu)=1-(1-\beta_\ell)^\nu$ is then
$$
\sum_{\nu\geq 0} e^{-\lambda_g}\frac{(\lambda_g)^\nu}{\nu!}c^{(\ell)}(\nu)
= 1-\exp(-\beta_\ell\lambda_g).
$$
Therefore, the four-class discrete-time mean-field equations are
$$\rho_{g,\ell}(t+1)=(1-\gamma)\rho_{g,\ell}(t)+[1-\rho_{g,\ell}(t)][1-\exp(-\beta_\ell\lambda_g)]$$
for $g,\ell\in\{1,2\}.$
At a positive fixed point,
\begin{equation*}\label{eq:fp}
    \gamma\rho_{g,\ell}^*=(1-\rho_{g,\ell}^*)[1-\exp(-\beta_\ell\lambda_g^*)].
\end{equation*}

These four equations must be solved self-consistently because $\lambda_1^*$ and $\lambda_2^*$ depend on the aggregate block prevalences [see Eqs.~\eqref{eq:agg_block_prevalence} and \eqref{eq:poisson_mean}].

Near the positive fixed point, 
\begin{align*}
    L_\nu^{(2)} &\propto \omega(1-\rho_{1,2}^*)\exp(-\lambda_1^*)\frac{(\lambda_1^*)^\nu}{\nu!} + (1-\omega)(1-\rho_{2,2}^*)\exp(-\lambda_2^*)\frac{(\lambda_2^*)^\nu}{\nu!},\\
    L_\nu^{(1)} &\propto (1-\omega)(1-\rho_{1,1}^*)\exp(-\lambda_1^*)\frac{(\lambda_1^*)^\nu}{\nu!} + \omega(1-\rho_{2,1}^*)\exp(-\lambda_2^*)\frac{(\lambda_2^*)^\nu}{\nu!}.
\end{align*}
Rather than working directly with $W_{\beta_2}(\nu)$ [Eq.~\eqref{eq:W2}], it is helpful to consider its odds,
\begin{equation}\label{eq:fp_W2}
    \frac{W_{\beta_2}(\nu)}{1-W_{\beta_2}(\nu)} 
    = \frac{L_\nu^{(2)}}{L_\nu^{(1)}}
    =\frac{\omega(1-\rho_{1,2}^*)+(1-\omega)(1-\rho_{2,2}^*)R(\nu)}{(1-\omega)(1-\rho_{1,1}^*)+\omega(1-\rho_{2,1}^*)R(\nu)},
\end{equation}
where 
$$
R(\nu)=\exp[-(\lambda_2^*-\lambda_1^*)]\left(\frac{\lambda_2^*}{\lambda_1^*}\right)^\nu
$$
is the ratio of the Poisson exposure probabilities for blocks 2 and 1.

\begin{figure}
    \centering
    \includegraphics[width=3.8in]{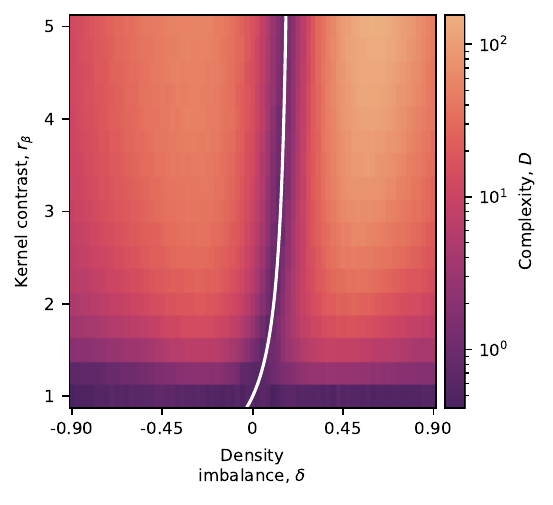}
    \caption{\textbf{The mean-field exposure-balance criterion predicts the location of the low-complexity region.}
    Complexity score, $D$, of the inferred global kernel mode with respect to density imbalance, $\delta$, and kernel contrast, $r_\beta\equiv\beta_2/\beta_1$ for the $(\varepsilon,\delta)$-SBM with $N=256$ nodes, $\langle k\rangle=18,$ $\varepsilon=0.9$, and $\omega=0.1$.
    We fix $\beta_1=0.01$ and vary $\beta_2\in[0.01, 0.05].$
    Each cell averages 10 simulations.
    The white curve shows the mean-field prediction obtained by solving the four-class mean-field system for its positive fixed point and locating the zero of $F(\delta)=\lambda_2^*(\delta)-\lambda_1^*(\delta),$ the value of $\delta$ at which the two structural blocks have equal mean exposure.}
    \label{fig:supplementary:mf_delta_curve}
\end{figure}

We see immediately that the quantity $W_{\beta_2}(\nu)$ is nearly constant when the two structural blocks generate similar exposure distributions. 
Under the Poisson approximation, this occurs when their mean exposure levels are approximately equal:
$$
\lambda_1^*(\delta) \approx \lambda_2^*(\delta),
$$
where the dependence of $\lambda_{g}$ on $\delta$ comes in through $p_{11}$ and $p_{22}$ in Eq.~\eqref{eq:poisson_mean}.
We therefore define
$$
F(\delta)=\lambda_2^*(\delta)-\lambda_1^*(\delta),
$$
where $\lambda_1^*$ and $\lambda_2^*$ are the block mean exposures [Eq.~\eqref{eq:poisson_mean}]  evaluated at the positive fixed point $(\rho_{1,1}^*,\rho_{1,2}^*,\rho_{2,1}^*,\rho_{2,2}^*)$ of the four-class mean-field system. 
The predicted low-complexity region is obtained by scanning over $\delta$---for fixed $N,\, \langle k\rangle,\, \varepsilon,\, \omega,\, \beta_1,$ and $\beta_2$ (equivalently, $r_\beta$), and $\gamma$---and identifying where $F(\delta)$ crosses zero.
The $\delta$-value at which $F(\delta)$ crosses zero is shown in \Cref{fig:supplementary:mf_delta_curve}.

This predicted zero is an implicit mean-field criterion for exposure balance: the value of $\delta$ at which the two structural blocks have approximately equal mean exposure levels.
At this point, the ratio $R(\nu)$ is approximately constant in $\nu$, so the quantity $W_{\beta_2}$ is also approximately constant in $\nu.$ 
The calculation should therefore be interpreted as a predictor for where exposure level is least informative about latent kernel class and thus where the global kernel is expected to have reduced apparent complexity.

\textbf{Reproducing experiments.}
For convenience, we collect the parameter values used in throughout our experiments.
All networks sampled from our $(\varepsilon,\delta)$-SBM model have $N=\NUMNODES$nodes and a mean degree $\langle k \rangle=\AVGDEGREE$.
We consider only unweighted, undirected networks without self-loops or multi-edges.
The block strength, $\varepsilon$, varies between experiments and takes values in $[0,1]$.

When we simulate contagion processes on these networks, we fix the healing rate at $\gamma \equiv 0.1$ and simulate all processes for \NUMSTEPS time steps. 
While the quality of the inference decreases with too few time steps, \NUMSTEPS time steps are more than sufficient for the size of networks considered here.
While the infection probability $\beta_1=0.01$ is fixed in all experiments, we vary $\beta_2$ to modify $r_\beta$.

\end{document}